\documentclass[twocolumn,showpacs,amsmath,amssymb,bibnotes]{revtex4}
\usepackage{bm}
\usepackage{graphicx,epsfig,subfigure,afterpage}
\usepackage{amsfonts}

\newcommand{\be}{\begin{equation}}
\newcommand{\ee}{\end{equation}}
\newcommand{\bea}{\begin{eqnarray}}
\newcommand{\eea}{\end{eqnarray}}
\newcommand{\gdot}{\dot{\gamma}}
\newcommand{\gdotbar}{\overline{\dot{\gamma}}}

\newcommand{\bw}{\begin{widetext}}
\newcommand{\ew}{\end{widetext}}
\newcommand{\lae}{\stackrel{<}{\scriptstyle\sim}}
\newcommand{\gae}{\stackrel{>}{\scriptstyle\sim}}

\newcommand{\xhat}{\vecv{\hat{x}}}

\newcommand{\zhat}{\vecv{\hat{z}}}

\newcommand{\vecv}[1]{\mathbf{{#1}}}
\newcommand{\tens}[1]{\mathbf{{#1}}}
\newcommand{\nablu}{{\bf \nabla}}

\begin{document}

\title{Vorticity structuring and Taylor-like velocity rolls triggered by gradient shear bands}
\author{Suzanne M. Fielding}
\email{suzanne.fielding@manchester.ac.uk}
\affiliation{School of Mathematics, University of Manchester, Booth
  Street East, Manchester M13 9EP, United Kingdom }

\date{\today}
\begin{abstract}

We suggest a novel mechanism by which vorticity structuring and
Taylor-like velocity rolls can form in complex fluids, triggered by
the linear instability of one dimensional gradient shear banded
flow. We support this with a numerical study of the diffusive Johnson-Segalman
model. In the steady vorticity structured state, the thickness of the
interface between the bands remains finite in the limit of zero stress
diffusivity, presenting a possible challenge to the accepted theory of
shear banding.

\end{abstract}
\maketitle


\section{Introduction}
\label{sec:intro}

\begin{figure}[tb]
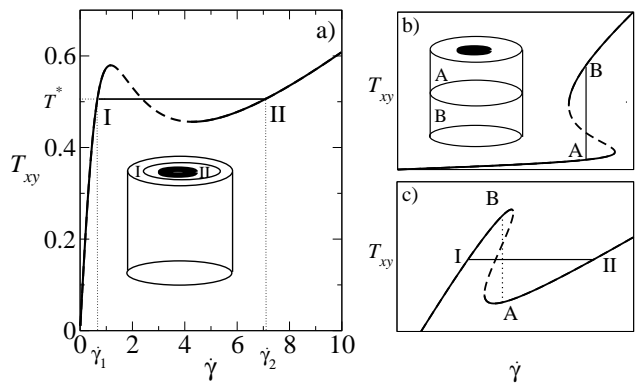

\subfigure{
  \includegraphics[scale=0.40]{./flowCurvea0.3.eps}
}
\hspace{-0.3cm}
\subfigure{
  \includegraphics[scale=0.35]{./schematic.eps}
}
\caption{a) Homogeneous constitutive curve and steady state flow curve for 1D planar gradient banded flow in the  DJS model at $a=0.3, \eta=0.05$. Inset: schematic arrangement of the bands in curved Couette flow. b) Schematic constitutive curve giving shear thickening vorticity banding. c) A shear thinning curve allowing gradient or vorticity banding in 1D. }
\label{fig:schematic}
\vspace{-0.5cm}
\end{figure}

Many complex fluids exhibit flow instabilities that result in
spatially heterogeneous, shear banded states.  Examples include
wormlike~\cite{BritCall97c,berret-pre-55-1668-1997,BecManCol04} and
onion~\cite{diat93,wilkinsOlms2006,SalManCol03} surfactants;
side-chain liquid crystalline polymers~\cite{pujolle01}; viral
suspensions~\cite{lettinga-jpm-16-S3929-2004,dhont-fd-123-157-2003};
telechelic polymers~\cite{berret-prl-8704--2001}; soft
glasses~\cite{CRBMGHJL02}; polymer solutions~\cite{HilVla02}; and
colloidal suspensions~\cite{ChenZAHSBG92}. In many cases, the
instability is explained by a region of negative slope
$dT_{xy}/d\gdot<0$ in the constitutive relation $T_{xy}(\gdot)$
between shear stress and shear rate for homogeneous flow, as shown in
Fig~\ref{fig:schematic}a. In this regime, homogeneous flow is unstable
~\cite{Yerushalmi70} with respect to the formation of bands of
differing shear rates $\gdot_1$ and $\gdot_2$, with layer-normals in
the flow-gradient direction.  Force balance requires that the shear
stress $T_{xy}$ is uniform across the gap, and therefore common to
both the bands. Any change in the overall applied shear rate
$\gdotbar$ causes a change in the relative volume fraction $f$ of the
bands according to a lever rule $\gdotbar=f\gdot_1+(1-f)\gdot_2$,
while $\gdot_1,\gdot_2$ and $T_{xy}$ remain constant. In bulk
rheometry, this leads to a plateau in the steady state flow curve at
some stress $T_{xy}=T^*$ (Fig.~\ref{fig:schematic}a), the value of
which is selected by accounting for spatial non-locality in the
constitutive dynamics of the viscoelastic
stress~\cite{lu-prl-84-642-2000}.  In what follows, we shall refer to
the effect just described as gradient shear banding, or simply
gradient banding.

In some systems, flow induced heterogeneity has been reported in the
vorticity direction. By analogy with the above discussion, this effect
is often termed vorticity banding. It has been observed in onion
surfactants~\cite{diat93,Bonn+98} and colloidal
crystals~\cite{ChenZAHSBG92}, accompanied by a steep stress ``cliff''
in the flow curve. Multiple turbid and clear vorticity bands also
occur in some polymeric~\cite{HilVla02} and
micellar~\cite{fischer-ra-41-35-2002,fischer-ra-39-234-2000}
solutions, accompanied by shear thickening. In these cases, the bands
not only alternate in space but also oscillate in time. In shear
thinning viral suspensions, multiple stationary vorticity bands can
arise in the regime of isotropic-nematic microphase
separation~\cite{lettinga-jpm-16-S3929-2004,dhont-fd-123-157-2003}.

In comparison with gradient banding, vorticity banding is poorly
understood theoretically.  To date, the main attempts to model it have
invoked the analogy with gradient
banding~\cite{goveas-epje-6-79-2001}. As discussed above, gradient
bands have different shear rates $\gdot_1,\gdot_2$, and coexist at a
common shear stress $T^*$. For vorticity bands, the moving rotor
imposes a shear rate that is common to each band. Pursuing the analogy
with gradient banding, it is natural to invoke an underlying
constitutive curve of the form in Fig.~\ref{fig:schematic}b, in the
case of shear thickening systems.  This allows bands (A and B) of
differing shear {\em stresses} to coexist, having layer normals in the
vorticity direction.  In steady state, one then expects a flow curve
with a steep stress cliff, consistent with the experimental
observations discussed above~\cite{diat93,Bonn+98,ChenZAHSBG92}. This
scenario can be adapted to shear thinning systems by invoking a
constitutive curve of the form sketched in Fig.~\ref{fig:schematic}c.
In fact, at the level of 1D calculations performed separately in the
flow-gradient and vorticity directions, such a curve can support
either gradient or vorticity banded
states~\cite{olmsted-pre-60-4397-1999}.  Which of these (if either)
would be selected in a full 3D calculation remains an outstanding
question.  Indeed, any concrete calculation of vorticity banding to
date has taken a simplified one-dimensional (1D) approach, allowing
spatial variations only in the direction of the layer normals, and
thereby imposing axial and radial symmetry.

Beyond the traditional shear banding literature, other flow
instabilities are well known to trigger structuring in the vorticity
direction, in both simple and complex fluids.  For simple liquids in
curved Couette flow, the unstable centripetal interplay of fluid
inertia with cell curvature gives rise to Taylor velocity
rolls~\cite{drazin} stacked in the vorticity direction. An analogous
inertia-free instability occurs in viscoelastic fluids, here triggered
by viscoelastic hoop stresses~\cite{larson-jfm-218-573-1990,Lars92c},
and again leading to velocity rolls stacked along the axis of the
Couette cylinder.  The rolls are typically observed as bandlike
structures, imaged by seeding the fluid with mica flakes.

In contrast to the 1D vorticity banding scenario discussed above, both
traditional and viscoelastic Taylor-Couette instabilities are
inherently 2D (at least), with velocity rolls comprising a circulation
of fluid in the flow-gradient/vorticity
plane~\cite{drazin,larson-jfm-218-573-1990}.  Not unexpectedly, given
this roll-like structure, the wavelength of the associated banding in
the vorticity direction is roughly set by the width of the gap in the
flow-gradient direction~\cite{drazin}.

In view of the above discussion, it is natural to ask whether any link
exists between traditional (1D) vorticity banding and 2D viscoelastic
Taylor-Couette instabilities; or whether the two effects are entirely
distinct. Given that both can be accompanied by shear thickening in
the bulk flow curve, this is a difficult question to address
experimentally. Even if they are distinct in theory, it seems feasible
that some experimental observations that have traditionally been
interpreted as 1D vorticity banding in fact comprise 2D viscoelastic
Taylor Couette rolls. Likely candidates include those systems in which
the wavelength of the alternating vorticity bands is comparable to the
width of the cell in the flow-gradient direction, suggesting an
underlying roll
structure~\cite{HilVla02,fischer-ra-41-35-2002,fischer-ra-39-234-2000,lettinga-jpm-16-S3929-2004,dhont-fd-123-157-2003}.
This was recently suggested in the context of viral suspensions in
Ref.~\cite{kang-pre-74--2006}. In other systems, particularly those
showing a very marked stress cliff in flow curve, the traditional 1D
scenario of Fig.~\ref{fig:schematic}b or c remains more likely.

In this paper, we suggest a novel mechanism by which vorticity
structuring can emerge in complex fluids. A key feature of our
approach is that, to some extent, it unifies traditional 1D (gradient)
banding descriptions with those of 2D roll-like instabilities. The
basic idea is as follows. The constitutive curve of
Fig.~\ref{fig:schematic}a gives rise initially to 1D gradient bands,
via the conventional instability in the region of negative slope,
$dT_{xy}/d\gdot<0$. These then undergo a secondary linear
instability~\cite{fielding-prl-95--2005,wilson-jnfm-138-181-2006}, due to the
action of normal stresses across the interface between the bands. This
leads finally to pronounced undulations along the interface, with
wavevector in the vorticity direction. These are accompanied by 2D
Taylor-like velocity rolls stacked in the vorticity direction, and
undulatory vorticity stress structuring superposed on the underlying
gradient bands.  In contrast to the conventional
inertial~\cite{drazin} and viscoelastic~\cite{larson-jfm-218-573-1990}
Taylor mechanisms, the vorticity instability introduced here does not
rely on cell curvature, but occurs even in the limit of planar shear,
to which our calculations are confined for simplicity.

The results to be presented are in good agreement with recent
experiments in which a gradient banded solution of wormlike micelles
was found to be unstable with respect to interfacial undulations with
wavevector in the vorticity direction~\cite{lerouge-prl-96--2006}. We
will return to a detailed comparison with these experiments later in
the manuscript.  Our results might also apply to systems in which
shear thickening~\cite{decruppe-pre-73--2006} and/or vorticity
banding~\cite{fischer-ra-41-35-2002,fischer-ra-39-234-2000,HilVla02,Bonn+98}
is reported to set in at the right hand edge of a stress plateau in
the flow curve (suggestive of underlying gradient banding, as in
Fig.~\ref{fig:schematic}a); or in which gradient and vorticity banding
have actually been observed
concomitantly~\cite{britton-prl-78-4930-1997}.

The paper is structured as follows. In Sec.~\ref{sec:model} we
introduce the diffusive Johnson Segalman model, within which the
subsequent calculations are to be performed. In Sec.~\ref{sec:1D} we
discuss 1D calculations, confined to the flow-gradient direction.
These predict gradient banding for applied shear rates in the regime
of negative slope in the homogeneous constitutive curve. In
Sec.~\ref{sec:linear} we switch to two dimensions -- the
flow-gradient/vorticity plane -- and show an initially 1D gradient
banded ``base state'' to be linearly unstable with respect to
undulations along the interface with wavevector in the vorticity
direction. We then perform a full 2D nonlinear numerical study of the
subsequent growth and eventual saturation of these undulations.
Details of the numerical method are discussed in
Sec.~\ref{sec:numerics}, followed by presentation of the results in
Sec.~\ref{sec:nonlinear}. Finally we summarise our findings and
discuss some directions for future study.

\section{Model and geometry}
\label{sec:model}

The generalised Navier--Stokes equation for a viscoelastic material in
a Newtonian solvent of viscosity $\eta$ and density $\rho$ is
\begin{equation} \label{eqn:NS} \rho(\partial_t +
  \vecv{v}.\nablu)\vecv{v} = \nablu .(\tens{\Sigma} + 2\eta\vecv{D}
  -P\tens{I}),
\end{equation} 
where $\vecv{v}(\vecv{r},t)$ is the velocity field and $\tens{D}$ is
the symmetric part of the velocity gradient tensor, $(\nablu
\vecv{v})_{\alpha\beta}\equiv
\partial_\alpha v_\beta$. The pressure field
$P(\vecv{r},t)$ is determined by enforcing incompressibility,
\be
\label{eqn:incomp}
\vecv{\nabla}\cdot\vecv{v}=0.
\ee
The quantity $\vecv{\Sigma}(\vecv{r},t)$ in Eqn.~\ref{eqn:NS} is the
extra stress contributed by the viscoelastic component. In principle,
the dynamics of this quantity should be explicitly derived by
averaging over the underlying microscopic dynamics of the viscoelastic
component. This was done in Refs.~\cite{cates90} for wormlike
micelles.  For simplicity, however, we use the phenomenological
Johnson-Segalman (JS) model~\cite{johnson-jnfm-2-255-1977}
\begin{widetext}
\be
\label{eqn:DJS}
(\partial_t
+\vecv{v}\cdot\nablu )\,\tens{\Sigma} 
= a(\tens{D}\cdot\tens{\Sigma}+\tens{\Sigma}\cdot\tens{D}) +
(\tens{\Sigma}\cdot\tens{\Omega} + \tens{\Omega}\cdot\tens{\Sigma})  
 + 2 G\tens{D}-\frac{\tens{\Sigma}}{\tau}+ \frac{\ell^2}{\tau }\nablu^2 
 \tens{\Sigma}.
\ee
\end{widetext}
In this equation, $G$ is a plateau modulus, $\tau$ is the viscoelastic
relaxation time, and $\tens{\Omega}$ is the antisymmetric part of the
velocity gradient tensor.  For $a=1$ and $\ell=0$, Eqn.~\ref{eqn:DJS}
reduces to the Oldroyd B model, which can be derived by considering
the dynamics of an ensemble of Hookean dumbbells in solution. For
$|a|<1$, the JS model captures non-affine slip between the dumbbells
and the solvent, leading to the drastic shear thinning of
Fig.~\ref{fig:schematic}a. Accordingly, $a$ is called the slip
parameter. The JS model is the simplest tensorial model to exhibit a
regime of negative slope in the homogeneous constitutive curve, and so
to predict a shear banding instability. As discussed further below,
the diffusive term $\nablu^2 \tens{\Sigma}$ in Eqn.~\ref{eqn:DJS} is
needed to correctly describe the ultimate shear banded
flow~\cite{lu-prl-84-642-2000}.

Within this model, we study planar shear between parallel plates at
$y=0,L$, with the top plate driven at velocity $V \xhat$.  At the
plates we assume boundary conditions of
$\partial_y\Sigma_{\alpha\beta}=0\;\forall\;\alpha,\beta$ for the
viscoelastic stress, with no slip and no permeation for the fluid
velocity.  In the linear stability analysis of Sec.~\ref{sec:linear},
we consider small values of the Reynolds number $Re=\rho L^2/\eta$. In
the nonlinear study of Secs.~\ref{sec:numerics}
and~\ref{sec:nonlinear}, we set $Re=0$ at the outset.  Throughout we
use units in which $G=1,\tau=1$ and $L=1$. 

\begin{figure}[b]
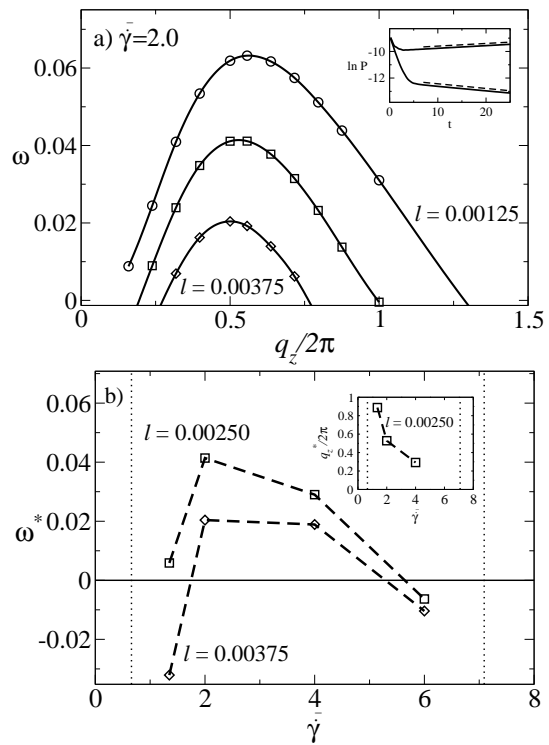

  \centering
\includegraphics[scale=0.4]{./dispersion.eps}
\includegraphics[scale=0.42]{./peak.eps}
 \caption{a) Dispersion relation for perturbations about a 1D banded
 state for $\ell=0.00375,\, 0.00250,\, 0.00125$. $a=0.3$, $\eta=0.05$,
 $\rho/\eta=0.02$. Inset: Linear dynamics of 2D code, starting from a
 flat interface. Solid lines: weight in modes, $q_zL_z/2\pi=1,2$;
 dashed: analytical prediction. b) Peak $\omega^*$ ($q^*$ in inset) in
 dispersion relation {\it vs.} $\gdotbar$ across the plateau of
 Fig.~\ref{fig:schematic}a. Vertical dashed lines denote the edges of
 the stress plateau. }
\label{fig:dispersion}
\vspace{-0.5cm}
\end{figure}

\section{1D gradient bands}
\label{sec:1D}

As noted above, to capture shear thinning the DJS model invokes a slip
parameter $a$ with $|a|<1$, giving non affine deformation of the
viscoelastic component~\cite{johnson-jnfm-2-255-1977}. The homogeneous
constitutive curve $T_{xy}=\Sigma_{xy}(\gdot, a)+\eta\gdot$ is then
capable of non-monotonicity, as in Fig.~\ref{fig:schematic}a.  For an
imposed shear rate $\gdotbar\equiv V/L$ in the region of decreasing
stress, homogeneous flow is unstable with respect to fluctuations with
wavevector in the flow-gradient direction
$y$~\cite{yerushal.j-ces-25-1891-1970}. A 1D calculation then predicts
separation into gradient bands of differing shear rates $\gdot_1,
\gdot_2$, with a flat interface in between. The diffusive term in
Eqn.~\ref{eqn:DJS} is needed to account for spatial gradients of the
shear rate and viscoelastic stress across the interface, which has a
characteristic thickness $O(\ell)$. It also ensures a unique,
history-independent banding stress
$T_{xy}=T^*$~\cite{lu-prl-84-642-2000}, as seen experimentally.  We
expect $\ell=O(10^{-4})$, set by the typical micellar mesh size, in
units of the (typical) gap size.


\section{Linear instability}
\label{sec:linear}

In Refs.~\cite{fielding-prl-95--2005,wilson-jnfm-138-181-2006}, we
considered the linear stability of this 1D gradient banded state with
respect to 3D ($x, y, z$) perturbations of infinitesimal amplitude. In
the flow direction $\xhat$ and vorticity direction $\zhat$ these are
decomposed into Fourier modes with wavevectors
$\vecv{q}=q_x\xhat+q_z\zhat$.  (Ref.~\cite{renardy} had previously
considered $\vecv{q}=q_x\xhat$ in the pathological limit $\ell=0$,
assuming ``top jumping''.)  For diffuse interfaces, $l\gae 0.015$, the
1D state is linearly stable. For $l\lae 0.015$, we find it to be
linearly unstable with respect to modes with wavevector
$\vecv{q}=q_x\xhat$ in the flow
direction~\cite{fielding-prl-95--2005,wilson-jnfm-138-181-2006}.  The
associated eigenfunction essentially corresponds to undulations along
the interface. For $l\lae 0.005$, the 1D state is also unstable with
respect to undulations with wavevector $\vecv{q}=q_z\zhat$ in the
vorticity direction. However, these modes are predicted to grow much
more slowly than those with $\vecv{q}=q_x\xhat$.  Accordingly, in
Refs.~\cite{fielding-prl-95--2005,wilson-jnfm-138-181-2006}, we
focused mainly on the dominant modes, $\vecv{q}=q_x\xhat$.
Subsequently in Ref.~\cite{fielding-prl-96--2006}, however, we showed
that these undulations are cut off, once they attain finite amplitude,
by the nonlinear effects of shear. In contrast, the vorticity
direction is neutral with respect to the shear. Accordingly, the modes
with $\vecv{q}=q_z\vecv{\hat{z}}$ should not suffer this cutoff and
are therefore likely to contribute significantly to the ultimate
nonlinear state, despite their much slower initial growth rate. With
this motivation, in this paper we study the dynamics of the model in
the flow-gradient/vorticity $(y-z)$ plane. For simplicity and
computational efficiency, we will assume uniformity in the flow
direction $x$, returning in Sec.~\ref{sec:nonlinear} below to comment
on the validity of this simplification.  It corresponds to taking a
vertical slice through one side of an axisymmetric flow state in the
planar limit of a Couette device.

The growth rates $\omega$ of the vorticity modes
$\vecv{q}=q_z\vecv{\hat{z}}$ are shown in Fig.~\ref{fig:dispersion}a,
at a single value of the imposed shear rate. States with thinner
interfaces (smaller $\ell$) are more unstable (larger $\omega>0$).
Fig.~\ref{fig:dispersion}b shows the growth rate of the maximally
unstable mode for shear rates across the stress plateau of
Fig.~\ref{fig:schematic}a. The corresponding wavelength
$\lambda^*=O(1)$ is of order the rheometer gap $L\equiv 1$ (inset). At
small $\ell$, the instability persists across most of the plateau, so
is likely to be unavoidable experimentally.  The mechanism of
instability is not fully understood, but is likely to stem from steep
gradients in the normal stress and shear rate across the
interface~\cite{hinch-jnfm-43-311-1992,mcleish87,fielding-prl-95--2005}.

\section{Numerical method}
\label{sec:numerics}

To study the undulations once they have grown to attain a finite
amplitude, beyond the regime of linear instability, we solve the
model's full nonlinear dynamics numerically.  In this section we
discuss the details of our numerical method, which is adapted from
that of Refs.~\cite{chilcott-jnfm-29-381-1988,fielding-prl-96--2006}.
Readers who are not interested in these issues can skip straight to
Sec.~\ref{sec:nonlinear} without loss of thread.

The model equations have already been specified in
Sec.~\ref{sec:model}, together with the flow geometry, boundary
conditions and choice of adimensionalisation. For computational
efficiency, our numerical study is confined to the $(y-z)$ plane,
assuming translational invariance along the flow direction $x$. In the
vorticity direction we take a cell of length $L_z$, with periodic
boundary conditions.

We consider the limit of zero Reynolds number, in which
Eqn.~\ref{eqn:NS} reduces to
\be
\label{eqn:zeroRey}
0 = \nablu .(\tens{\Sigma} + 2\eta\vecv{D} -P\tens{I}).
\ee
To ensure that the incompressibility constraint of
Eqn.~\ref{eqn:incomp} is satisfied always, we express the velocity in terms
of stream functions $\phi$ and $\psi$:
\be
\label{eqn:stream}
v_x=\partial_y\phi,\;\; v_y=\partial_z\psi,\;\; v_z=-\partial_y\psi.
\ee
In this way, Eqn.~\ref{eqn:incomp} need no longer be considered: it
remains only solve Eqns.~\ref{eqn:DJS} and~\ref{eqn:zeroRey}, with the
velocity expressed as in Eqn.~\ref{eqn:stream}.

To solve these, the basic strategy is to step along a grid of time
values $t^n=n\Delta t$ for $n=1,2,3\cdots$, at each step updating
$\tens{\Sigma}^n, \phi^n, \psi^n\to \tens{\Sigma}^{n+1}, \phi^{n+1},
\psi^{n+1}$.  Discretization with respect to time of any quantity $f$
is denoted $f(t^n)=f^n$, or sometimes below by $f|^n$. At each
time-step, we first update the viscoelastic stress $\tens{\Sigma}^n \to
\tens{\Sigma}^{n+1}$ using the constitutive equation (\ref{eqn:DJS})
with fixed, old values of the stream-functions $\phi^n,\psi^n$. We
then update $\phi^n,\psi^n \to \phi^{n+1},\psi^{n+1}$ using the force
balance equation (\ref{eqn:zeroRey}) with the new values of
$\tens{\Sigma}^{n+1}$.

The update $\tens{\Sigma}^n\to\tens{\Sigma}^{n+1}$ using the
viscoelastic constitutive equation (\ref{eqn:DJS}) is performed as
follows. As a preliminary step, we rewrite (\ref{eqn:DJS}) in the form
\be
\label{eqn:DJSsep}
\partial_t\tens{\Sigma}= 
 f(\nablu \vecv{v},\tens{\Sigma}) - \vecv{v}\cdot\nablu \tens{\Sigma}
 + \ell^2 \nablu^2 \tens{\Sigma}, 
\ee
in which $f(\nablu \vecv{v},\tens{\Sigma})$ comprises the
non-diffusive terms from the right hand side of Eqn.~\ref{eqn:DJS}.
In what follows, the three terms on the right hand side of
Eqn.~\ref{eqn:DJSsep} are referred to as the local, advective and
diffusive terms respectively. Numerically, they are dealt with in
three successive partial updates $\tens{\Sigma}^n \to
\tens{\Sigma}^{n+1/3}$, $\tens{\Sigma}^{n+1/3} \to
\tens{\Sigma}^{n+2/3}$ and $\tens{\Sigma}^{n+2/3} \to
\tens{\Sigma}^{n+1}$.

In the first of these, the local term is handled using an explicit
Euler algorithm~\cite{PreTeuVetFla92}, checked for consistency against
a fourth order Runge-Kutta algorithm~\cite{PreTeuVetFla92}.
Temporarily setting aside the issue of spatial discretization, the
Euler algorithm can be written
\be
\label{eqn:local}
\tens{\Sigma}^{n+1/3}(y,z)=\tens{\Sigma}^n + \Delta t \; f(\nablu
\vecv{v}^n,\tens{\Sigma}^n).
\ee
In terms of the stream-functions $\phi$ and $\psi$, the
velocity-gradient tensor $\nablu \vecv{v}$ has Cartesian components
\be
\label{eqn:localM}
\nablu \vecv{v}= \left(\begin{array}{ccc}
            0 & 0 & 0 \\
            \partial_y^2\phi & \partial_y\partial_z\psi &
            -\partial_y^2\psi \\
            \partial_y\partial_z\phi & \partial_z^2\psi &
            -\partial_y\partial_z\psi \end{array}\right),
\ee
in which we have omitted the superscripts $n$ for clarity.
Eqns.~\ref{eqn:local} and~\ref{eqn:localM} are then spatially
discretized on a rectangular grid in real space. In some runs of the
code, the grid points are linearly spaced, with $z_i=i\Delta z$ for
$i=1\cdots N_z$ and $y_j=j\Delta y$ for $j=1\cdots N_y$.  In others,
we used a nonlinear mapping in the $y$ direction to focus attention on
the region explored by the interface. For simplicity, most of the
description of this section will concern the linear grid, though we
will return briefly at the end of the section to discuss the nonlinear
modification.  In either case, any spatially discretized function $f$
is denoted $f(y^j,z^i)=f_{ij}$, or sometimes $f|_{ij}$.  (The
apparently unusual order of the indices is a historical convention on
the part of the author, stemming from a previous study in the $x-y$
plane.)  Eqn.~\ref{eqn:local} then becomes
\begin{widetext}
\be
\label{eqn:localD}
\tens{\Sigma}^{n+1/3}_{ij}=\tens{\Sigma}^n_{ij} + \Delta t \; f(\nablu
\vecv{v}^n_{ij},\tens{\Sigma}^n_{ij}),
\ee
The derivatives in the components of $\nablu\vecv{v}^n_{ij}$ are
discretized (in the case of a rectangular grid) as follows:
\be
\label{eqn:p2y}
\partial_y^2\psi|^n_{ij}=\frac{1}{ \Delta
  y^2}\left[\psi_{i(j+1)}^n-2\psi_{ij}^n+\psi_{i(j-1)}^n\right],
\ee
\be
\partial_z^2\psi|^n_{ij}=\frac{1}{ \Delta
  z^2}\left[\psi_{(i+1)j}^n-2\psi_{ij}^n+\psi_{(i-1)j^n}\right],
\ee
and
\be
\partial_y\partial_z \psi|^n_{ij}=\frac{1}{4\Delta x\Delta
  y}\left[\psi_{(i+1)(j+1)}-\psi_{(i+1)(j-1)}-\psi_{(i-1)(j+1)}+\psi_{(i-1)(j-1)}\right].
\ee
Corresponding derivatives of $\phi$ are obtained in the same way,
replacing $\psi$ by $\phi$ in the above equations. For values of $ij$
at the edges of the flow domain, these formulae link to values of the
flow variables at ``phantom'' grid points that lie just outside the
domain.  These values are specified by imposing the boundary
conditions, the spatial discretization of which is discussed at the
end of this section.

The advective term is also handled using an explicit Euler
algorithm~\cite{PreTeuVetFla92}, on the same real space grid:
\bea
\label{eqn:advect}
\tens{\Sigma}^{n+2/3}_{ij}&=&\tens{\Sigma}^{n+1/3}_{ij} - \Delta t
\left(
v_{y}|_{ij}^n \, \partial_y\tens{\Sigma}|^n_{ij}+v_{z}|_{ij}^n\,\partial_z\tens{\Sigma}|^n_{ij}\right)\nonumber\\
  &=&\tens{\Sigma}^{n+1/3}_{ij} - \Delta t
\left(
\partial_z\psi |_{ij}^n \, \partial_y\tens{\Sigma}|^n_{ij}-\partial_y\psi|_{ij}^n\,\partial_z\tens{\Sigma}|^n_{ij}\right).
\eea
The derivatives of $\psi$ in this equation are discretized as follows:
\be
\label{eqn:p1y}
\partial_y\psi|_{ij}^n=\frac{1}{2\Delta
  y}\left[\psi_{i(j+1)}-\psi_{i(j-1)}\right]\;\;\;\textrm{and}\;\;\; \partial_z\psi|_{ij}^n=\frac{1}{2\Delta
  z}\left[\psi_{(i+1)j}-\psi_{(i-1)j}\right].
\ee
The derivative of $\tens{\Sigma}$ with respect to $y$ in
Eqn.~\ref{eqn:advect} was discretized using third-order
upwinding~\cite{Pozrikidis}:
\be
\partial_y\tens{\Sigma}_{ij}^n=\frac{1}{6\Delta
  y}\left[\tens{\Sigma}^n_{i(j-2)} -6 \tens{\Sigma}^n_{i(j-1)} +3
  \tens{\Sigma}^n_{ij} +2 \tens{\Sigma}^n_{i(j+1)}
\right]\;\;\;\textrm{if}\;\;\; v_y|^n_{ij}>0,
\ee
while
\be
\partial_y\tens{\Sigma}_{ij}^n=\frac{1}{6\Delta
  y}\left[-\tens{\Sigma}^n_{i(j+2)} +6 \tens{\Sigma}^n_{i(j+1)} -3
  \tens{\Sigma}^n_{ij} -2 \tens{\Sigma}^n_{i(j-1)}
\right]\;\;\;\textrm{if}\;\;\; v_y|^n_{ij}<0,
\ee
with analogous expressions for the derivative of $\tens{\Sigma}$ with
respect to $z$.

The diffusive term is handled by discretizing on $y$ in real space as
above, taking a fast Fourier transform $z\to q_i$ in the vorticity
dimension using a standard NAG routine~\cite{NAG}, and solving the
resulting problem using a semi-implicit Crank-Nicolson
algorithm~\cite{PreTeuVetFla92}:
\be
\label{eqn:Crank}
\tens{\Sigma}^{n+1}_{ij}-\tens{\Sigma}^{n+2/3}_{ij} =\frac{1}{2} l^2
\Delta t
\left(\partial_y^2\tens{\Sigma}_{ij}^{n+2/3}-q_i^2\tens{\Sigma}_{ij}^{n+2/3}\right)+\frac{1}{2}
l^2 \Delta t
\left(\partial_y^2\tens{\Sigma}_{ij}^{n+1}-q_i^2\tens{\Sigma}_{ij}^{n+1}\right),
\ee
in which the index $i$ now labels the Fourier mode number. The
derivatives $\partial_y^2$ are discretized as in Eqn.~\ref{eqn:p2y}
above. Note that Eqn.~\ref{eqn:Crank} contains no mixing of the
Cartesian components $\Sigma_{\alpha\beta}$ for any
$\alpha\beta=xx,xy,yy\cdots$, so can be solved for each one
separately. Bringing all terms in the unknown
$\tens{\Sigma}^{n+1}_{ij}$ across to the left hand side, and putting
all terms in the known $\tens{\Sigma}^{n+2/3}_{ij}$ on the right hand
side, we obtain a sparse set of linear equations characterized by a
tridiagonal matrix on the left hand side. These are then solved for
the $\tens{\Sigma}^{n+1}_{ij}$ using standard NAG routines~\cite{NAG}.

Having updated $\tens{\Sigma}^n \to \tens{\Sigma}^{n+1}$ using the
viscoelastic constitutive equation, we now update the stream-functions
$\phi^n,\psi^n \to \phi^{n+1},\psi^{n+1}$ using the $x,y$ and $z$
components of the force balance equation (\ref{eqn:zeroRey}). Again,
we work in real flow-gradient space and reciprocal vorticity space. To
eliminate the pressure from Eqn.~\ref{eqn:zeroRey}, we subtract
$\partial_y$ of the $z$ component from $\partial_z$ of the $y$
component to get the following equations, written separately for
$q_i=0$ and $q_i\neq 0$:
\be
\label{eqn:fb1}
\partial_y^3\phi|^{n+1}_{0(j+1/2)}=-\frac{1}{\eta}\partial_y\Sigma_{xy}|_{0(j+1/2)}^{n+1}\;\;\;\textrm{for}\;\;\; q_i=0,
\ee
\be
\label{eqn:fb2}
\partial^3_y\phi|^{n+1}_{i(j+1/2)}-q_i^2 \partial_y\phi|^{n+1}_{i(j+1/2)}=-\frac{1}{\eta}\left(\partial_y\Sigma_{xy}|^{n+1}_{i(j+1/2)}  + iq_i \Sigma_{xz}|^{n+1}_{i(j+1/2)}\right)\;\;\;\textrm{for}\;\;\; q_i\neq 0,
\ee
\be
\label{eqn:fb3}
\partial_y^3\psi|^{n+1}_{0(j+1/2)}=\frac{1}{\eta}\partial_y\Sigma_{yz}|^{n+1}_{0(j+1/2)}\;\;\;\textrm{for}\;\;\;q_i=0,
\ee
\be
\label{eqn:fb4}
\nablu^4\psi|^{n+1}_{ij}=-\frac{1}{\eta}\left[iq_i\partial_y\left(\Sigma_{yy}-\Sigma_{zz}\right)|^{n+1}_{ij}-\left(q_i^2+\partial_y^2\right)\Sigma_{yz}|^{n+1}_{ij}\right]\;\;\;\textrm{for}\;\;\;
q_i\neq
0,
\ee
with $\nablu^2=(\partial_y^2-q_i^2)$. The real and imaginary parts of
these equations are treated separately. The third order equations
(\ref{eqn:fb1}), (\ref{eqn:fb2}) and (\ref{eqn:fb3}) are discretized
at staggered half grid points $y_{j+1/2}$ for $j=1\cdots N_y-1$, with
derivatives calculated as follows:
\be
\partial_y f|_{j+1/2}=\frac{1}{\Delta y}\left(f_{j+1}-f_j\right),
\ee
and 
\be
\partial_y^3 f|_{j+1/2}=\frac{1}{\Delta
  y^3}\left(f_{j+2}-3f_{j+1}+3f_{j}-f_{j-1}\right),
\ee
for any quantity $f$. (For clarity, the subscript $i$ and the
superscript $n$ have been omitted from these expressions.)  The fourth
order equation is implemented at full grid points $y_j$, excluding
those at the very edge of the domain ($y_0$ and $y_N$). In it,
$\partial^2_y$ and $\partial_y$ are discretized as in
Eqns.~\ref{eqn:p2y} and~\ref{eqn:p1y} respectively, and $\partial_y^4$
according to
\be
\partial_y^4 f|_{j}=\frac{1}{\Delta y^4}\left(f_{j+2}-4f_{j+1}+6f_j
  -4f_{j-1}+f_{j-2}\right).
\ee
Each of Eqns.~\ref{eqn:fb1} to~\ref{eqn:fb4}, for each mode index $i$,
then takes the form of a sparse set of linear equations for
$\phi_{ij}^{n+1}$ or $\psi_{ij}^{n+1}$. These equations are solved
using standard NAG routines~\cite{NAG}.

\end{widetext}

It remains finally to specify the spatial discretization of the
boundary conditions. In turn, this will prescribe the values of the
flow variables on the phantom grid points that lie just outside the
flow domain.

In the vorticity direction $z$, the boundary conditions are periodic.
For any quantity $f$ on the grid $z_1\cdots z_{N_Z}$ in real space
$z_i$, we thus have $f_{-1}=f_{{N_z}-1}$, $f_{0}=f_{{N_Z}}$, $f_{{N_z}+1}=f_0$,
$f_{{N_z}+2}=f_1$. In reciprocal space $q_i$, the periodic boundary conditions
are always satisfied.

In the flow-gradient direction $y$, the boundary conditions for the
fluid velocity at the plates $y=0,1$ are those of no slip and no
permeation. In terms of the stream-function $\phi$, the no slip
condition gives $\partial_y\phi=0$ at $y=0$ and
$\partial_y\phi=\gdotbar$ at $y=1$. We also note that $\phi$ is only
defined up to an arbitrary additive constant, and accordingly choose
$\phi=0$ at $y=0$. The third order Eqns.~\ref{eqn:fb1}
and~\ref{eqn:fb2} then have three boundary conditions, as required.
After discretizing real flow-gradient space $y_j$ in reciprocal
vorticity space $q_i$, we then have
$$\phi_{i1}=0,\;\;\;\phi_{i0}=\phi_{i2},\;\;\;\textrm{and}\;\;\;\phi_{i({N_y}+1)}=\phi_{i({N_y}-1)}+\delta_{i0}\gdotbar \Delta y,$$
in which $\delta_{ij}$ is the usual Kronecker delta function.

In terms of the stream-function $\psi$, the no-slip condition gives
$\partial_y\psi=0$ at $y=0,1$, and the no-permeation condition gives
$\partial_z\psi=0$ at $y=0,1$. We also note that $\psi$ is only
defined up to an arbitrary additive constant, and choose $\psi=0$ at
$y=0$. In the $q_i=0$ equation (\ref{eqn:fb3}), we then have
$$\psi_{i1}=0,\;\;\;\psi_{i0}=\psi_{i2}\;\;\;\textrm{and}\;\;\;\psi_{i({N_y}+1)}=\psi_{i({N_y}-1)}.$$
These also hold for the $q_i\neq 0$ equation (\ref{eqn:fb4}), which
obeys the additional condition
$$\psi_{i{N_y}}=0.$$
The zero-gradient boundary condition for the viscoelastic stress, after
discretization on the flow-gradient grid $y_1,y_2\cdots y_{N_y}$, gives
$f_{i0}=f_{i2}$ and $f_{i({N_y}+1)}=f_{i({N_y}-1)}$ for all components
$f=\Sigma_{xx}, \Sigma_{xy}, \Sigma_{yy}\cdots$. These apply in both
real $z_i$ and reciprocal $q_i$ vorticity spaces.

Extension to the case of a nonlinear grid in the $y$ direction is
straightforward in principle, but cumbersome in detail. Discretized
derivatives are calculated via the usual Taylor expansions. For
example, to first order accuracy, centred second derivatives with
respect to $y$ become
\be
\partial_y^2
f_j=\frac{2}{y_{j+1}-y_{j-1}}\left[\frac{f_{j+1}-f_j}{y_{j+1}-y_j}-\frac{f_j-f_{j-1}}{y_j-y_{j-1}}\right].
\ee
We checked our nonlinear mapping carefully by performing a few runs
for identical parameter sets with both linear and highly nonlinear
grids.  All the numerical results in this paper are converged with
respect to grid and timestep, to within the accuracy resolvable on the
plots presented.

\begin{figure}[t]
\vspace{-0.8cm}\includegraphics[scale=0.63,
angle=90]{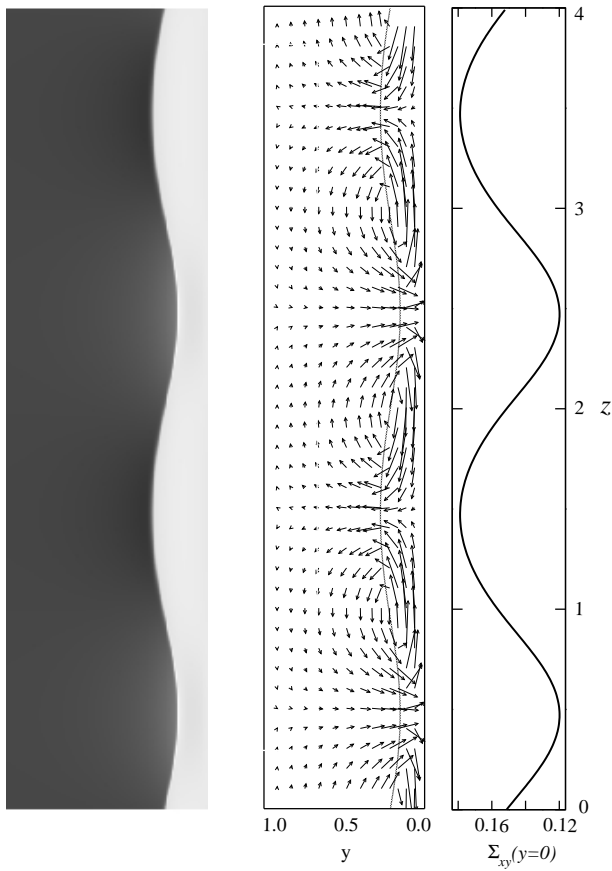}
\includegraphics[scale=0.44, angle=90]{./flowMap.eps}
\caption{Steady state at $a=0.3$, $\eta=0.05$, $\gdotbar=2.0$, $\ell=0.00375$, $L_z=4.0$. Left: greyscale of $\Sigma_{xx}$ in $y-z$ plane. Middle: $(y-z)$ velocity vectors, showing Taylor-like rolls. Right: vorticity banding of viscoelastic shear stress.}
\label{fig:steady}
\vspace{-0.5cm}
\end{figure}

\section{Nonlinear steady state}
\label{sec:nonlinear}

In each simulation run, we input as an initial condition the 1D
gradient-banded state discussed in Sec.~\ref{sec:1D}, superposed with
Fourier perturbations of tiny random amplitudes.  As expected, under
conditions where the linear analysis of Sec.~\ref{sec:linear} predicts
the 1D initial state to be unstable with respect to perturbations with
wavevectors $\vecv{q}=q_z\zhat$ in the vorticity direction, we find
that these initial disturbances grow in time. Full agreement between
(i) the early-time growth rate and functional form of the fastest
growing mode and (ii) the most unstable eigenvalue and eigenfunction
of the linear stability analysis provides a stringent check of our
numerical method.

During the instability, the initially flat interface between the bands
develops undulations that grow in time.  At long times, once nonlinear
effects become important, these saturate in a finite amplitude
interfacial undulation to give a 2D steady state
(Fig.~\ref{fig:steady}).  The wavelength of the steady undulations
corresponds to that of the maximally unstable mode of the linear
analysis. For the parameters of Fig.~\ref{fig:steady}, this is roughly
twice the gap width.  Associated with these undulations are
Taylor-like velocity rolls stacked in the vorticity direction
(Fig.~\ref{fig:steady}, middle), accompanied by undulations of the
stress along the wall (Fig.~\ref{fig:steady}, right).  The results of
Fig.~\ref{fig:steady} could be tested experimentally as follows.
Optical measurements should reveal birefringent stripes stacked in the
vorticity direction, each of height comparable to the gap width.
Likewise, the velocity rolls could potentially be measured using
velocimetry. This is a challenging task, however, because the highest
speed in Fig.~\ref{fig:steady}c is only $O(0.01)$.

Our results capture recent experimental observations in which an
initially 1D gradient banded state of a wormlike surfactant solution
was found to destabilise with respect to interfacial undulations with
wavevector in the vorticity direction~\cite{lerouge-prl-96--2006}.
Indeed, several features of our results can be directly compared with
these experiments, as follows. In the ultimate steady state, the
wavelength $O(L)$ and amplitude $O(L/10)$ of the undulations in our
Fig.~\ref{fig:steady} are comparable to those in Fig.  2 of
Ref.~\cite{lerouge-prl-96--2006}, measured in units of the gap width
$L$.  The wavelength of these undulations was furthermore reported to
increase with increasing average applied shear rate
$\gdotbar$~\cite{lerouge-prl-96--2006}, consistent with the inset of
our Fig.~\ref{fig:dispersion}b.
\begin{figure}[t]
  \centering
\includegraphics[scale=0.42]{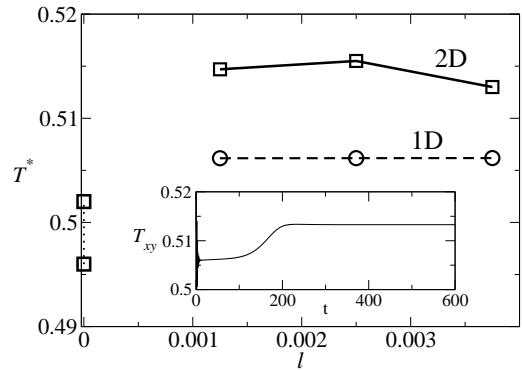}
 \caption{Selected stress: 1D initial state and 2D steady
 state. $L_z=2.0$, $a=0.3$, $\eta=0.05$, $\gdotbar=2.0$. At $\ell=0$, the
 stress varies erratically about an average that depends on the
 grid. Inset: evolution to steady state at $l=0.00375$. }
\label{fig:selected}
\vspace{-0.5cm}
\end{figure}

The kinetics of the instability can also be compared, via the temporal
evolution of the stress signal. In the
experiments~\cite{lerouge-prl-96--2006}, a shear startup protocol was
followed. Accordingly, the stress signal showed an initial overshoot
followed by a decay (at $\gdotbar=30s^{-1}$) on a timescale $O(\tau)$
to a plateau value. This part of the dynamics corresponded to the
initial formation of 1D gradient bands.  It is absent from our
simulations, because we take as our initial condition an already 1D
gradient banded flow.  Subsequently, the stress signal in
Ref.~\cite{lerouge-prl-96--2006} slowly increased by about $1\%$ on a
timescale $O(100\tau)$.  This part of the dynamics was associated with
the 1D gradient banded state destabilising to exhibit vorticity
undulations. As shown in Fig.~\ref{fig:selected}, it is captured very
well by our simulations: we find a slow stress increase $O(1\%)$ on a
time-scale $O(100\tau)$, consistent with the experiments.

Some differences between our work and the experiments of
Ref.~\cite{lerouge-prl-96--2006} are noted as follows.  In
Ref.~\cite{lerouge-prl-96--2006}, the instability was studied using
light scattering techniques, which couple to concentration
fluctuations. In the present manuscript, we do not consider
concentration coupling. In future work, it might be interesting to
perform analogous simulations in the concentration coupled model of
Ref.~\cite{fielding-epje-11-65-2003}.  However, an important finding
of the present work is that concentration coupling is not actually
needed to trigger the basic undulatory instability.  Indeed, we
believe this to stem instead from normal stress effects, with
concentration coupling a sub-dominant feature. Finally, we have not
seen the exotic dynamics reported at the edges of the stress plateau
in Ref.~\cite{lerouge-prl-96--2006}. We cannot access small enough
$\ell$ for the instability to persist here.  (Recall
Fig.~\ref{fig:dispersion}b.)

\begin{figure}[t]
\subfigure{
  \includegraphics[scale=0.47]{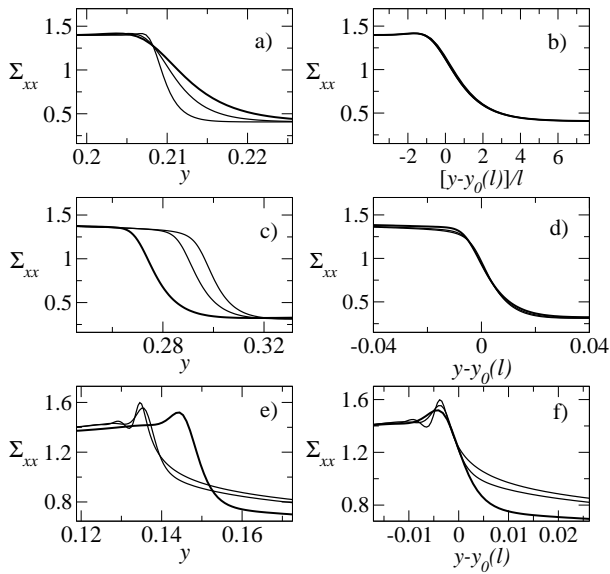}
}
\caption{Profile $\Sigma_{xx}$ normal to the interface; $\ell=0.00375$ (thick lines), $\ell=0.00250, 0.00125$ (thin). $a=0.3$, $\eta=0.05$, $\gdotbar=2.0$, $L_z=2.0$. a,b) 1D state with interfacial thickness $d\propto \ell$. (c,d), (e,f) 2D  state at the $z-$coordinate where the interface has maximal $+$ and $-$ $y$ displacements. The  thickness $d$ appears {\em independent} of $\ell$, but note the  bump of thickness $O(l)$. The offset $y_0(\ell)$ is chosen to centre the profile at the origin in each case.}
\label{fig:interface}
\vspace{-0.5cm}
\end{figure}

We return to comment on the validity of restricting our study to the
$y-z$ plane, which was done mainly for computational efficiency. As
seen from our results, the ultimate amplitude of the interfacial
undulation in this ($y-z$) plane is in fact comparable to that
reported in the ($x-y$) plane in
Ref.~\cite{fielding-prl-96--2006} (to within $10\%$ at $a=0.3$,
$\eta=0.05$, $\gdotbar=2.0$, $\ell=0.00375$, $L_{\{x,z\}}=2$). The
present study therefore shows that vorticity structuring is indeed
important, but also that a full 3D simulation should be performed in
future work.

Finally, we discuss briefly the thickness $d$ of the interface between
the bands.  In the 1D initial state, $d=O(\ell)$. In the limit
$\ell\to 0$, this gives an unphysically sharp interface $d\to
0$. Associated with this is a pathological steady state that strongly
depends strongly on the flow history~\cite{olmsted-jr-44-257-2000}. In
2D, in contrast, $d$ appears virtually independent of $\ell$, as shown
in Fig.~\ref{fig:interface}c-f.  This is an important finding that
could potentially obviate the gradient term $l^2\nabla^2\tens\Sigma$
in Eqn.~\ref{eqn:DJS}, which is needed to give a finite interfacial
thickness in 1D~\cite{lu-prl-84-642-2000}.  Nonetheless, the
interfacial profile does retain a small bump of thickness $O(\ell)$,
Fig.~\ref{fig:interface}e-f, suggesting that the local case $\ell=0$
remains pathological even in 2D. Indeed, at $\ell=0$ the stress signal
varies erratically about an average that varies between runs,
Fig.~\ref{fig:selected}, though purely numerical instability cannot be
ruled out. This important issue will be pursued further in future
work.

To summarise, we have identified a mechanism by which vorticity stress
bands and Taylor-like velocity rolls can form in a complex fluid,
triggered by the instability of gradient shear banded flow with
respect to interfacial undulations. In any real startup experiment, we
would expect the vorticity instability to commence during the final
stages of the initial band formation: above we assumed a complete
separation of timescales between the processes. In future work, we
will study the true dynamics of shear startup experiment in curved
Couette flow. We will also extend to 3D, to study the interplay of
vorticity banding with the dynamics of
Ref.~\cite{fielding-prl-96--2006}. Robustness of the mechanism in
other models will also be studied.

 The author thanks Peter Olmsted, Paul Callaghan, Georgina Wilkins and
 Helen Wilson for discussions, and the UK's EPSRC for financial
 support, GR/S29560/01.


\begin{thebibliography}{10}

\bibitem{BritCall97c}
M.~M. Britton and P.~T. Callaghan, Phys. Rev. Lett. {\bf 78},  4930  (1997).

\bibitem{berret-pre-55-1668-1997}
J.~F. Berret, G. Porte, and J.~P. Decruppe, Phys. Rev. E {\bf 55},  1668
  (1997).

\bibitem{BecManCol04}
L. Becu, S. Manneville, and A. Colin, Phys.\ Rev.\ Lett. {\bf 93},  art. no.
  (2004).

\bibitem{diat93}
O. Diat, D. Roux, and F. Nallet, J.~Phys.~II (France) {\bf 3},  1427  (1993).

\bibitem{wilkinsOlms2006}
G.~M.~H. Wilkins and P.~D. Olmsted, Submitted for publication  (2006).

\bibitem{SalManCol03}
J.~B. Salmon, S. Manneville, and A. Colin, Phys.\ Rev.\ E {\bf 68},  art. no.
  (2003).

\bibitem{pujolle01}
C. Pujolle-Robic and L. Noirez, Nature {\bf 409},  167  (2001).

\bibitem{lettinga-jpm-16-S3929-2004}
M.~P. Lettinga and J.~K.~G. Dhont, J. Physics-condensed Matter {\bf 16},  S3929
   (2004).

\bibitem{dhont-fd-123-157-2003}
J.~K.~G. Dhont, M.~P. Lettinga, Z. Dogic, T.~A.~J. Lenstra, H. Wang, S.
  Rathgeber, P. Carletto, L. Willner, H. Frielinghaus, and P. Lindner, Faraday
  Discussions {\bf 123},  157  (2003).

\bibitem{berret-prl-8704--2001}
J.~F. Berret and Y. Serero, Phys. Rev. Lett. {\bf 8704},    (2001).

\bibitem{CRBMGHJL02}
P. Coussot, J.~S. Raynaud, F. Bertrand, P. Moucheront, J.~P. Guilbaud, H.~T.
  Huynh, S. Jarny, and D. Lesueur, Phys.\ Rev.\ Lett. {\bf 88},  art. no.
  (2002).

\bibitem{HilVla02}
L. Hilliou and D. Vlassopoulos, Ind.\ Eng.\ Chem.\ Res. {\bf 41},  6246
  (2002).

\bibitem{ChenZAHSBG92}
L.~B. Chen, C.~F. Zukoski, B.~J. Ackerson, H.~J.~M. Hanley, G.~C. Straty, J.
  Barker, and C.~J. Glinka, Phys. Rev. Lett. {\bf 69},  688  (1992).

\bibitem{Yerushalmi70}
J. Yerushalmi, S. Katz, and R. Shinnar, Chemical Engineering Science {\bf 25},
  1891  (1970).

\bibitem{lu-prl-84-642-2000}
C.~Y.~D. Lu, P.~D. Olmsted, and R.~C. Ball, Phys. Rev. Lett. {\bf 84},  642
  (2000).

\bibitem{Bonn+98}
D. Bonn, J. Meunier, O. Greffier, A. Alkahwaji, and H. Kellay, Phys. Rev. {\bf
  E} {\bf 58},  2115  (1998).

\bibitem{fischer-ra-41-35-2002}
P. Fischer, E.~K. Wheeler, and G.~G. Fuller, Rheologica Acta {\bf 41},  35
  (2002).

\bibitem{fischer-ra-39-234-2000}
P. Fischer, Rheologica Acta {\bf 39},  234  (2000).

\bibitem{goveas-epje-6-79-2001}
J.~L. Goveas and P.~D. Olmsted, European Phys. J. E {\bf 6},  79  (2001).

\bibitem{olmsted-pre-60-4397-1999}
P.~D. Olmsted and C.~Y.~D. Lu, Phys. Rev. E {\bf 60},  4397  (1999).

\bibitem{drazin}
P.~G. Drazin and W.~H. Reid, {\em Hydrodynamic stability} (Cambridge University
  Press, Cambridge, 2004).

\bibitem{larson-jfm-218-573-1990}
R.~G. Larson, E.~S.~G. Shaqfeh, and S.~J. Muller, J. Fluid Mechanics {\bf 218},
   573  (1990).

\bibitem{Lars92c}
R.~G. Larson, Rheol. Acta. {\bf 31},  213  (1992).

\bibitem{kang-pre-74--2006}
K.~G. Kang, M.~P. Lettinga, Z. Dogic, and J.~K.~G. Dhont, Phys. Rev. E {\bf
  74},    (2006).

\bibitem{fielding-prl-95--2005}
S.~M. Fielding, Phys. Rev. Lett. {\bf 95},    (2005).

\bibitem{wilson-jnfm-138-181-2006}
H.~J. Wilson and S.~M. Fielding, J. Non-newtonian Fluid Mechanics {\bf 138},
  181  (2006).

\bibitem{lerouge-prl-96--2006}
S. Lerouge, M. Argentina, and J.~P. Decruppe, Phys. Rev. Lett. {\bf 96},
  (2006).

\bibitem{decruppe-pre-73--2006}
J.~P. Decruppe, O. Greffier, S. Manneville, and S. Lerouge, Phys. Rev. E {\bf
  73},    (2006).

\bibitem{britton-prl-78-4930-1997}
M.~M. Britton and P.~T. Callaghan, Phys. Rev. Lett. {\bf 78},  4930  (1997).

\bibitem{cates90}
M.~E. Cates, J.~Phys. Chem. {\bf 94},  371  (1990).

\bibitem{johnson-jnfm-2-255-1977}
M.~W. Johnson and D. Segalman, J. Non-newtonian Fluid Mechanics {\bf 2},  255
  (1977).

\bibitem{yerushal.j-ces-25-1891-1970}
W. Yerushal.j, S. Katz, and R. Shinnar, Chem. Engineering Science {\bf 25},
  1891  (1970).

\bibitem{renardy}
Y.~Y. Renardy, The. Comp. Fl.~Dyn. {\bf 7},  463  (1995).

\bibitem{fielding-prl-96--2006}
S.~M. Fielding and P.~D. Olmsted, Phys. Rev. Lett. {\bf 96},    (2006).

\bibitem{hinch-jnfm-43-311-1992}
E.~J. Hinch, O.~J. Harris, and J.~M. Rallison, J. Non-newtonian Fluid Mechanics
  {\bf 43},  311  (1992).

\bibitem{mcleish87}
T.~C.~B. McLeish, J.~Poly. Sci. B-Poly. Phys. {\bf 25},  2253  (1987).

\bibitem{chilcott-jnfm-29-381-1988}
M.~D. Chilcott and J.~M. Rallison, J. Non-newtonian Fluid Mechanics {\bf 29},
  381  (1988).

\bibitem{PreTeuVetFla92}
W.~H. Press, S.~A. Teukolsky, W.~T. Vetterling, and B.~P. Flannery, {\em
  Numerical Recipes in C (2nd ed.)} (Cambridge University Press, Cambridge,
  1992).

\bibitem{Pozrikidis}
C. Pozrikidis, {\em Introduction to Theoretical and Computation Fluid Dynamics}
  (Oxford University Press, New York, 1997).

\bibitem{NAG}
Numerical Algorithms Group Ltd., Wilkinson House, Jordan Hill Road, Oxford, OX2
  8DR, UK.

\bibitem{fielding-epje-11-65-2003}
S.~M. Fielding and P.~D. Olmsted, European Phys. J. E {\bf 11},  65  (2003).

\bibitem{olmsted-jr-44-257-2000}
P.~D. Olmsted, O. Radulescu, and C.~Y.~D. Lu, J. Rheology {\bf 44},  257
  (2000).

\end{thebibliography}


\end{document}